\documentclass[preprintnumbers,amsmath,amssymb,showpacs,twocolumn]{revtex4}

\usepackage{graphicx}
\usepackage{dcolumn}
\usepackage{bm}

\begin{document}

\title{ Two avowable quantum communication schemes }

\author{Feng-Li Yan$^1,$ Dong Ding$^{1,2}$}
\affiliation {$^1$ College of Physics  Science and Information
Engineering, Hebei Normal
University, Shijiazhuang 050016, China\\
$^2$ North China Institute of Science and Technology, Langfang
101601,  China }

\date{\today}

\begin{abstract}

  Two avowable quantum communication schemes are proposed.
  One is an avowable teleportation protocol  based on the quantum
  cryptography. In this protocol one teleports  a set of one-particle states based on the
  availability of an honest arbitrator, the keys and the  Einstein-Podolsky-Rosen  pairs
  shared by the communication parties and the arbitrator. The key point is that
  the fact of the teleportation can neither  be disavowed by the sender nor  be
  denied by the receiver. Another  is an avowable quantum secure direct
  communication  scheme. A one-way Hash function chosen by the
  communication parties helps the receiver to validate the truth of
  the information and to avoid disavowing for the sender.
  \end{abstract}


\maketitle

\section{Introduction}

One of the essential features of quantum information is its capacity
for entanglement.  Entanglement is a uniquely quantum mechanical
resource that plays a key role in many of the most interesting
applications of quantum computation and quantum information, such as
quantum teleportation \cite {s1,s12,s13,s14,s2}, quantum key
distribution \cite {s3,s4,s5,WangPr07,GRTZrmp02}, quantum secure
direct communication \cite {s6,s7,s8,s9,LongGuiLuzongshu}, quantum
secret sharing  \cite {HBB99,s11} and so on.

When  the pure state entanglement is shared by the sender Alice and
the receiver Bob, it allows them to send quantum data with classical
communication via teleportation \cite {s1}. Quantum computation and
quantum information have revealed a plethora of methods for
interchanging resources, many built upon quantum teleportation, so
it is commonly understood as one of the most important aspects of
quantum information theory. Quantum teleportation has attracted
widespread attention since the seminal work on teleportation by
Bennett et al \cite {s1}.  So far research work on quantum
teleportation  has got great development, theoretical and
experimental \cite {s12,s13,s14,s2} as well.

Another important aspect of quantum information is quantum secure
direct communication, in which the two parties communicate important
messages directly without first establishing a shared secret key to
encrypt them and the message is deterministically sent through the
quantum channels, but can be read only after obtaining an additional
classical information for each bit \cite {s6,s7,s8,s9}.

The purpose of  classical signature is to  guarantee the
communication process neither  to be disavowed by the sender nor  to
be denied by the receiver. In other words,   the signature can
ensure the validity of the communication. Recently, Zeng et al \cite
{s15} and Gottesman et al \cite {s16} proposed the quantum signature
schemes by combining the classical signature idea and quantum
cryptography. Obviously the avowable (or signatory) communication
schemes  are needed in the
 modern society. In present paper we will first propose  an avowable (or signatory) teleportation protocol
based on the
  availability of an honest arbitrator, the keys and the  Einstein-Podolsky-Rosen (EPR) pairs
  shared by the communication parties and the arbitrator.
    Then  an avowable scheme of
quantum secure direct communication based on digital signature will
be presented. Here the receiver can validate the truth of the
information and avoid disavowing for the sender with the virtue of a
one-way Hash function chosen by the
  communication parties.

\section{An avowable  teleportation scheme based on quantum cryptograph}

Quantum teleportation is a technique for moving quantum states
around, even in the absence of a quantum communication channel
linking the sender of the quantum state to the recipient.  An
avowable teleportation scheme means  that  the quantum teleportation
can neither  be disavowed by the sender nor  be denied by the
receiver.

Assume that there are members Alice, Bob and an honest arbitrator
Charlie in a communication group. Now the sender Alice wants to
transmit a set of unknown single-particle states
\begin{equation}
|u\rangle_i=\alpha_i|0\rangle_i+\beta_i|1\rangle_i,
\\
|\alpha_i|^2+|\beta_i|^2=1, i=1,2,\cdots, n,
\end{equation}
to the receiver Bob. In order to realize avowable teleportation we
need an honest arbitrator Charlie sharing the secret keys $K_a$ and
$K_b$ with Alice and Bob respectively. Of course, the secret keys
$K_a$ and $K_b$ can be generated via mature quantum cryptography,
for example BB84 protocol \cite {s3}, so that the keys are
unconditional security \cite {s17}. We also suppose that the
arbitrator can  make and distribute EPR pairs
\begin{equation}|\phi^+\rangle_{C_AA}=\frac
{1}{\sqrt 2}(|00\rangle+|11\rangle)_{C_AA}
\end{equation}
between him and Alice, and
\begin{equation}|\phi^+\rangle_{C_BB}=\frac {1}{\sqrt
2}(|00\rangle+|11\rangle)_{C_BB}
\end{equation}
between him and  Bob. Here $C_A$ and $C_B$ indicate the particles
held by Charlie. Now we present the avowable teleportation scheme in
details as follows.

1) Alice sends the arbitrator an application indicating that she
wants to teleport a set of  unknown quantum states to the receiver
Bob. The communication between Alice and Charlie should be encrypted
by using the key $K_a$.

2) When the arbitrator Charlie has  received the encrypted message
of the application, he  decrypts the secret information to prove
Alice's identity. After that the EPR pair channel between Alice and
Bob is established by using entanglement swapping
\cite{Entanglementswapping}.
  That is, Charlie makes
a measurement on the particles $C_A, C_B$ in the Bell base
$\{|\phi^+\rangle_{C_A C_B},
 |\phi^-\rangle_{C_A C_B},
|\psi^+\rangle_{C_A C_B}, |\psi^-\rangle_{C_A C_B}\},$ where
\begin{eqnarray}
\nonumber&&|\phi^+\rangle_{C_A C_B}=\frac {1}{\sqrt
2}(|00\rangle+|11\rangle)_{C_A C_B},\\\nonumber
 &&|\phi^-\rangle_{C_A C_B}=\frac
{1}{\sqrt 2}(|00\rangle-|11\rangle)_{C_A C_B},\\\nonumber
&&|\psi^+\rangle_{C_A C_B}=\frac {1}{\sqrt
2}(|01\rangle+|10\rangle)_{C_A C_B},\\ &&|\psi^-\rangle_{C_A
C_B}=\frac {1}{\sqrt 2}(|01\rangle-|10\rangle)_{C_A C_B}.
\end{eqnarray}
As
\begin{eqnarray}
\nonumber&&|\phi^+\rangle_{C_AA}|\phi^+\rangle_{C_BB}
\\\nonumber=&&\frac
{1}{\sqrt 2}(|00\rangle+|11\rangle)_{C_AA}\frac {1}{\sqrt
2}(|00\rangle+|11\rangle)_{C_BB}\\\nonumber =&&\frac
{1}{2}(|\phi^+\rangle_{C_AC_B}|\phi^+\rangle_{AB}+|\phi^-\rangle_{C_AC_B}|\phi^-\rangle_{AB}\\
&&+|\psi^+\rangle_{C_AC_B}|\psi^+\rangle_{AB}+|\psi^-\rangle_{C_AC_B}|\psi^-\rangle_{AB}),\end{eqnarray}
so the outcome of the Bell measurement on the particles $C_A,C_B$
determines the state of the particles  $A,B$ \cite
{Entanglementswapping}. Then Charlie sends $n$ outcomes of the Bell
 measurement on the particles $C_A, C_B$ to Alice secretly by
using the key $K_a$.

3) When Alice has received the secret results of the measurement
sent by Charlie, she decrypts it. If the result of the measurement
is $|\phi^+\rangle$ ($|\phi^-\rangle$, $|\psi^+\rangle$,
$|\psi^-\rangle$), Alice performs a unitary transformations $I$ (
$\sigma_z$, $\sigma_x$, $i\sigma_y$) on the particle $A$ to change
the state of the particles $A$ and $B$ into
\begin{equation}
|\phi^+\rangle_{AB}=\frac {1}{\sqrt 2}(|00\rangle+|11\rangle)_{AB}.
\end{equation}
Here $I$ is the identity operator, $\sigma_x,\sigma_y,\sigma_z$ are
Pauli operators. By now $n$ perfect EPR pairs have been built
between Alice and Bob.

4) Alice makes a Bell  measurement on the particles $i$ and $A$. The
results of the Bell measurement are written as
$\{|Bell\rangle_{iA}\}$. Then she encrypts them using $K_a$ (the key
between Alice and Charlie) to get
\begin{equation}
S_a=K_a(\{|Bell\rangle_{iA}\}).
\end{equation}
After that she informs Charlie about the outcome of $S_a$ via the
classical channels.

As a matter of fact, the  overall  state of the particles $i$, $A$,
and $B$ can be written as
\begin{eqnarray}\nonumber
&&|\Psi\rangle_{iAB}\\\nonumber
 =&&|u\rangle_i\otimes |\phi^+\rangle_{AB}\\\nonumber
 =&&\frac
{1}{\sqrt
2}(\alpha_i|000\rangle+\beta_i|100\rangle+\alpha_i|011\rangle
+\beta_i|111\rangle)_{iAB}\\\nonumber
 =&&\frac
{1}{2}[|\phi^{+}\rangle_{iA}(\alpha_i|0\rangle+
\beta_i|1\rangle)_B+|\phi^{-}\rangle_{iA}(\alpha_i|0\rangle-
\beta_i|1\rangle)_B\\\nonumber
 &&+|\psi^+\rangle_{iA}(\beta_i|0\rangle+
\alpha_i|1\rangle)_B+|\psi^-\rangle_{12}(-\beta_i|0\rangle+
\alpha_i|1\rangle)_B].\\\nonumber\\
\end{eqnarray}
Here $|\phi^{\pm}\rangle_{iA}=\frac {1}{\sqrt 2}(|00\rangle\pm
|11\rangle)_{iA}$ and $|\psi^\pm\rangle_{iA}=\frac {1}{\sqrt
2}(|01\rangle\pm |10\rangle)_{iA}$ are four Bell states of particles
$i$ and $A$. Thus when   Alice makes a Bell measurement on two
particles $i, A$, then regardless of the identity of $|u\rangle_i$,
each outcome will occur with equal probability $\frac {1}{4}$. Hence
after this measurement
 the resulting state of Bob's particle will
be respectively
\begin{eqnarray}
&&(\alpha_i|0\rangle+ \beta_i|1\rangle)_B=I |u\rangle_B,\\\nonumber
&&(\alpha_i|0\rangle-
\beta_i|1\rangle)_B=\sigma_z|u\rangle_B,\\\nonumber
&&(\beta_i|0\rangle+
\alpha_i|1\rangle)_B=\sigma_x|u\rangle_B,\\\nonumber
&&(-\beta_i|0\rangle+ \alpha_i|1\rangle)_B=i\sigma_y |u\rangle_B.
\end{eqnarray}
Obviously, in each case  the state of Bob's particle is related to
$|u\rangle$  by a fixed unitary transformation independent of the
identity of $|u\rangle$. Thus if Bob obtains Alice's actual Bell
measurement outcome, then Bob will be able to apply the
corresponding inverse unitary transformation to particle $B$,
restoring it to state $|u\rangle_B$ in each case.

5) Charlie decrypts $S_a$ and obtains $\{|Bell\rangle_{iA}\}$ using
the key $K_a$. Then he encrypts $\{|Bell\rangle_{iA}\}$ using the
key $K_b$ (the key between Bob and Charlie) to obtain
\begin{equation}
S_c=K_b(\{|Bell\rangle_{1A}(i)\}).
\end{equation}
Later on  Charlie sends  message $S_c$ to Bob via the classical
channel.

6) Bob decrypts $S_c$ and obtains $\{|Bell\rangle_{iA}\}$
 using the key $K_b$, then Bob can perform a series of appropriate
 unitary transformations according to the information of
 $\{|Bell\rangle_{iA}\}$, respectively. By far, quantum states
 $\{ |u\rangle_i\}$, $(i=1,2,\ldots,n)$ has been teleported from the sender Alice to the receiver Bob successfully.

 Avowable is the notable feature of this teleportation protocol.
 Since  Alice's key $K_a$ and Bob's key $K_b$ are contained  in the
 classical information $S_a$ and $S_c$, if Alice disavows her
 behavior, it is very easy to be discovered. At the same time, Bob
 can not deny having received the quantum state also. Hence the process of above teleportation is of legalization.

\section{An avowable  quantum secure direct communication scheme based on digital signature}

In this section we propose an avowable quantum secure direct
communication scheme based on digital signature. A one-way Hash
function chosen by the
  communication parties helps the receiver to validate the truth of
  the information and to avoid disavowing for the sender.

Suppose that Alice  would like to transmit a message $M$,  a
classical bit string $M=\{m_1,m_2,\ldots,m_n\}$,  to Bob secretly,
where $m_k\in \{0,1\}$. We also assume that there are two channels
between Alice and Bob. One is a classical communication channel;
 the other is a quantum
communication channel, which is assumed to be insecure and the
eavesdropper can manipulate the quantum signal in any way she
desires. Next we introduce a procedure of the avowable quantum
secure direct communication scheme based on digital signature. The
steps of the protocol are as follows.

1) Alice and Bob choose a one-way Hash function  $H(x)$ together.
The  one-way Hash function implies  that it is  easy to calculate
$H(M)$ from $M$ but it is very difficult to calculate $M$ from
$H(M)$. Here the Hash function can be obtained through a faithful
arbitrator, he will be  a notarial center when the sender is
disavowable.

2) Alice prepares $l$ EPR pairs  in the state $\frac {1}{\sqrt
2}(|00\rangle+|11\rangle)_{AB}$. The set of these $l$ EPR pairs  is
  expressed as $\{\phi_{AB}(j)\}$ $(j=1,2,\ldots,l)$, where $l\approx 2n$, then she sends  particle $B$ of each EPR
  pair to Bob via the quantum channel.

 3) After having received $l$ particles  sent by Alice,  Bob choose randomly
  $(l-n)\approx n$ particles to test the security of the quantum channel.
  He measures these particles in the  basis
  $\{|0\rangle, |1\rangle\}$ or  $\{|+\rangle, |-\rangle\}$ at random, where
  $|\pm\rangle=\frac {1}{\sqrt 2}(|0\rangle\pm |1\rangle)$,   then he
  tells Alice the  base sequence he used. Alice chooses the same bases to perform measurements
  on  the corresponding  particles in her side and announces the measurement results. Then they
  compare their results publicly via the classical channel, if the results are indeed
  perfectly correlated, it is shown that the channels is secure.
  We express  the  untested subset of the EPR pairs  as
  $\{\phi_{AB}(i)\}$ $(i=1,2,\ldots,n)$, and  Alice's and Bob's
  particles  as $\{\phi_{A}(i)\}$ $(i=1,2,\ldots,n)$
  and $\{\phi_{B}(i)\}$ $(i=1,2,\ldots,n)$, respectively. So far a
  perfect EPR pair channel has be established between Alice and Bob.

4) Alice encrypts the message $M$ on $n$ EPR pairs. If $m_k$ is 0,
then Alice performs a unitary transformation $I=|0\rangle\langle
0|+|1\rangle\langle 1|$ on the particle $\phi_{A}(k)$; if $m_k$ is
1, she applies a unitary transformation $\sigma_x=|0\rangle\langle
1|+|1\rangle\langle 0|$, one of the Pauli operators, on the particle
$\phi_{A}(k)$. After performing the unitary transformation the
resulting state of the EPR pair will be
\begin{equation}
I|\phi\rangle_{AB}=\frac {1}{\sqrt 2}(|00\rangle+|11\rangle)_{AB}
\end{equation}
and
\begin{equation} \sigma_x|\phi\rangle_{AB}=\frac {1}{\sqrt
2}(|10\rangle+|01\rangle)_{AB},
\end{equation}
respectively.

 5)   Alice performs $\{|0\rangle, |1\rangle\}$ base
measurements on $\{\phi_{A}(i)\}$ and announces the results of her
measurements via the classical channel.

6) Alice transforms $M$ into $H(M)$ by the Hash function which they
had chosen together, then she sends $H(M)$ to Bob publicly.

7) After Bob  received the outcomes of Alice's measurement and
$H(M)$, Bob performs $\{|0\rangle, |1\rangle\}$ base measurements on
particles $\{\phi_{B}(i)\}$. By using equations (11) and (12) we can
conclude that if  the  outcome of  Alice's measurement  is the same
with that of Bob's, the message $m_k$ is 0; otherwise the message
$m_k$ is 1. Hence according to the results of both Alice's and Bob's
measurement, Bob can obtain the message $M'$.

8) Bob calculates  $H(M')$ from  $M'$  by the same Hash function
which they had chosen together, and compares $H(M)$ and $H(M')$, if
$H(M)=H(M')$, the signature of the message $M$ is true, then accepts
it, otherwise discards it.

In our scheme, two parties of the communication, Alice and Bob,
transform the information freely, until the sender is disavowable.
Because the common Hash function is obtained through an arbitrator,
who can testify whether the sender tell a lie or not. At the same
time, this scheme can validate the truth of the information, if the
message is not authentic, then $H(M)\neq H(M')$, the receiver Bob
will discard it.

\section{Security}

Our schemes are  secure. It is easy to see that these two schemes
can not only prevent Eve from attacking but also avoid disavowing.
The former is a quantum teleportation based on the quantum
cryptography, where we choose the method of BB84 protocol that is
unconditionally secure  \cite {s17}, to generate and distribute the
keys as well as to detect the eavesdropping of Eve. The scheme gives
a new method for teleportation, in which  the fact of the
teleportation can neither be disavowed by the sender nor  be
  denied by the receiver.
Because there are personal keys $K_a$ and $K_b$ within the classical
information $S_a$ and $S_c$, the disavowable behavior must be
discovered by the faithful arbitrator. The latter is a quantum
secure direct communication scheme based on digital signature. Since
$H(x)$ is a one-way function, hence it is difficult for Eve to  read
the message $M$ from $H(M)$.  The test performed by Alice and Bob
can guarantee  the security of the EPR pairs which will be used to
transmit the secret message.  Since the channel is safe, when Bob
gets $H(M)=H(M')$, he can conclude that the signature of message $M$
is true. At the same time, a one-way Hash function that Alice and
Bob possessed together is distributed by the faithful arbitrator, so
if the sender is disavowable, whose action can be found easily.

\section {Conclusion}

In conclusion, we have proposed two avowable  protocols for quantum
communication. One is a teleportation protocol based on the quantum
cryptography, which  can avoid disavowing. Another is a scheme for
quantum secure direct communication based on digital signature, in
which the receiver can validate the truth of the information and
avoid disavowing for the sender. It is shown that our schemes are
 secure.\\

{\noindent\bf Acknowledgments}\\[0.2cm]

 This work was supported by the National Natural Science Foundation of China under Grant No: 10671054, Hebei Natural Science Foundation of China under Grant
Nos: A2005000140, 07M006, and  the Key Project of Science and
Technology Research of Education Ministry of China under Grant No:
207011.

\end{document}